\documentstyle[twocolumn,prb,aps,epsfig,graphics]{revtex}

\begin{document}
\draft
\twocolumn[\hsize\textwidth\columnwidth\hsize\csname
@twocolumnfalse\endcsname 
\title{Polaron effects in electron channels on a helium film}
\author{G. A. Farias and R. N. Costa Filho}
\address{Departamento de F\'{\i}sica, Universidade Federal do Cear\'{a},\\
Campus do Pici, Caixa Postal 6030, 60455-760 Fortaleza,\\
Cear\'{a}, Brazil}
\author{F. M. Peeters}
\address{Universiteit Antwerpen (UIA), Departement Natuurkunde,\\
Universiteitsplein 1, B-2610 Antwerpen, Belgium}
\author{Nelson Studart}
\address{Departamento de F\'{\i}sica Universidade Federal de S\~{a}o Carlos,%
\\
Caixa Postal 676,  13565-905, S\~{a}o Carlos, S\~{a}o Paulo, Brazil}
\maketitle

\begin{abstract}
Using the Feynman path-integral formalism we study the polaron effects in
quantum wires above a liquid helium film. The electron interacts with
two-dimensional (2D) surface phonons, i.e. ripplons, and is confined in one
dimension (1D) by an harmonic potential. The obtained results are valid for
arbitrary temperature ($T$), electron-phonon coupling strength ($\alpha $),
and lateral confinement ($\omega _{0}$). Analytical and numerical results
are obtained for limiting cases of $T$, $\alpha $, and $\omega _{0}$. We found
the surprising result that reducing the electron motion from 2D to quasi-1D
makes the self-trapping transition more continuous.
\end{abstract}

\pacs{63.20.Kr, 71.38.+i} 
\vskip2pc] 
\section{INTRODUCTION}

In recent years multi-channel electron systems have been realized over
liquid helium by using one-dimensional (1D) metallic gate structures or by
providing a surface distortion through a corrugated substrate in addition to
the holding electric field perpendicular to the surface.\cite{rev97} A
single channel was also obtained using a polymer groove\cite{kiri93} and
strips on a circuit board.\cite{annamaria98} In these quasi-one-dimensional
(Q1D) systems, the electron motion is quantized in the $z$ direction by the
effective holding potential due to an external electric field and image
forces coming from the substrate, and in the $y$ direction by a lateral
confinement potential which can be modelled quite accurately, as in a wide
class of Q1D and Q0D electron systems, by a parabolic well.\cite{sokolov95}

It is well known that the interaction of the electron with the quantized
surface excitations (ripplons, in the case of a liquid surface) can be
described as a polaron problem with the amazing possibility of a
phase-transition like behavior between localized and delocalized states when
varying the electron-ripplon coupling.\cite{polaron} The ripplonic polaron
state on the surface of helium results into a decrease of the conductivity
of the electron layer. More detailed information of the polaron state can be
obtained by measuring the electron conductivity as a function of the
microwave frequency.\cite{dahm97} Even though the formation of the localized
polaron state has been claimed, there is no conclusive interpretation about
the origin of the mobility dip in some experiments and the question is still
controversial.\cite{controversial}

Quite recently 1D confinement effects on surface electrons on bulk helium
have been addressed by using the hydrodynamic model of the polaron which
describes the energetics of the dimple formation on the surface of liquid
helium and its transport properties.\cite{sokolov00}

Even though surface excitations of the liquid helium film have in general a
complicated dispersion relation coming from contributions of surface
tension, gravity and film thickness, in the case of thin films the ripplon
spectrum has a well-defined acoustical character.\cite{polaron} In a
previous work,\cite{peeters85} the 3D acoustical polaron was studied within
the Feynman approach and it was found that, as a function of the
electron-phonon coupling constant $\alpha $, the polaron undergoes a
self-trapping transition at $\alpha \sim \sqrt{k_{0}},$ where $k_{0}$ is a
finite Debye cutoff in phonon space, which is continuous for $k_{0}<18\ $
and becomes discontinuous when $k_{0}>18$. For a 2D system, a similar
self-trapping transition was found at $\alpha \simeq 0.25$ which is almost $%
k_{0}$-independent. The transition is now continuous (discontinuous) for $%
k_{0}<130$ $(k_{0}>130).$\cite{farias96}

In this paper, we will investigate how the polaron effect is altered when we
further confine the electron system to a channel, i.e. a Q1D system. In the
case of LO-phonon interaction it is well known that the polaron effect
(renormalization) increases with confinement, i.e. with reduction of the
dimensionality.\cite{peeters86,gil90} Here, we will study the competition
between confinement and polaron effects as resulting from the interaction
with acoustical phonons (ripplons). The confinement alters the available
phase space of allowed $k$-vectors for virtual phonon emission and
absorption and in this way the polaron effect.

The present paper is organized as follows. In Sec. II, we present the Fr\"{o}%
hlich-type Hamiltonian, the path-integral Feynman formalism,\cite%
{feynman55,books} the trial Hamiltonian used and the variational principle
to derive an upper bound to the exact free energy of the system studied.
Explicit analytic results in limiting cases of temperature $%
T$, electron-phonon coupling strength $\alpha $, and lateral
confinement $\omega _{0}$ are presented in Sec. III. The numerical results
are shown in Sec. IV. In the last section we will present our conclusions.

\section{HAMILTONIAN MODEL AND FREE ENERGY}

The Hamiltonian which describes an electron interacting with 2D phonons and
subject to a lateral confinement due to an harmonic potential is given by 
\begin{equation}
H={\frac{{\vec{p}}^2}{2m}}+{\frac 12}m\omega _0^2y^2+\sum_{\vec{k}}\hbar
\omega _{\vec{k}}(a_{\vec{k}}^{\dagger }a_{\vec{k}}+{\frac{{}_1}{{}^2}})+H_I,
\label{H}
\end{equation}
with the electron-phonon interaction given by 
\begin{equation}
H_I=\sum_{\vec{k}}\left( V_{\vec{k}}a_{\vec{k}}e^{i\vec{k}\cdot \vec{r}}+V_{%
\vec{k}}^{*}a_{\vec{k}}^{\dagger }e^{-i\vec{k}\cdot \vec{r}}\right) ,
\label{HI}
\end{equation}
where $\vec{r}$ $(\vec{p})$ is the 2D position (momentum) operator of the
electron with mass $m$, $V_{\vec{k}}$ is the electron-phonon strength and $%
a_{\vec{k}}^{\dagger }$ $(a_{\vec{k}})$ are the creation (annihilation)
operators of the phonons with wave vector $\vec{k}$, and frequency $\omega _{%
\vec{k}}$. The electron is laterally confined by an harmonic potential with
frequency $\omega _0$.

The Helmholtz free energy is given by

\begin{equation}
F=-{\frac{1}{\beta }}\log Z,  \label{Freal}
\end{equation}
with

\begin{equation}
Z=Tre^{-\beta H},  \label{Zreal}
\end{equation}
the partition function. The free energy, Eq. (\ref{Freal}), and the
partition function, Eq. (\ref{Zreal}), contain the contribution of the free
phonons, $F_{ph}$. Here, we will shift the free energy $F$ by this constant
contribution $F_{ph}$. Using the path-integral formalism, the trace of the
partition function \cite{feynman55,osaka,schultz} can be written as

\begin{equation}
Z=\int d{\vec{r}}\int D{\vec{r}}(u)\exp {\{S[{\vec{r}}(u)]\}}\delta ({\vec{r}%
}(\beta )-{\vec{r}})\delta ({\vec{r}}(0)-{\vec{r}}),  \label{Zfey}
\end{equation}
where $\int D{\vec{r}}$ denotes the integral over all possible electron
paths, $\beta =1/k_{B}T$ is the inverse temperature, and $u$ is related to
the real time variable $t$ as $u=it$. The action $S[{\vec{r}}]$ in Eq. (\ref%
{Zfey}) is obtained after the exact elimination of the phonon coordinates 
and is given by \cite{osaka}

\begin{mathletters}
\begin{equation}
S=S_e+S_I,  \label{Sreal}
\end{equation}
with

\begin{equation}
S_{e}=-{\frac{m}{{2}}}\int_{0}^{\beta }dt({\dot{x}(t)}^{2}+{\dot{y}(t)}^{2})-%
{\frac{1}{{2}}}m{\omega _{0}}^{2}\int_{0}^{\beta }dty(t)^{2},  \label{Sereal}
\end{equation}

\begin{equation}
S_I=\sum_{\vec{k}}{|V_{\vec{k}}|}^2\int_0^\beta dt\int_0^\beta dsG_{\omega
_k}(t-s)e^{\imath {\vec{k}}\cdot (\vec{r}(t)-\vec{r}(s))},  \label{SIreal}
\end{equation}
where

\begin{equation}
G_\omega (t)={\frac 12}[(1+n(\omega ))e^{-\hbar \omega \mid t\mid
}++n(\omega )e^{\hbar \omega \mid t\mid }]  \label{Gomega}
\end{equation}
is the phonon Green's function, and $n(\omega )=(e^{\beta \hbar \omega
}-1)^{-1}$ the number of phonons with frequency $\omega $.

The path-integral in Eq. (\ref{Zfey}), can not be evaluated exactly. We
follow Feynman's polaron theory\cite{feynman55} and introduce a trial action 
$S_{0}$, such that: i) the path integral with this action can be done
exactly, and (ii) it approximates the original action $S$ as close as
possible. The free energy associated with this trial action is given by

\end{mathletters}
\begin{eqnarray}
Z_0=e^{-\beta F_0}=&&\int d{\vec{r}}\int D{\vec{r}}(u)\exp {\{S_0[{\vec{r}}%
(u)]\}}\times  \nonumber \\
&&\delta ({\vec{r}}(\beta )-{\vec{r}})\delta ({\vec{r}}(0)-{\vec{r}}),
\label{Ztrial}
\end{eqnarray}
and we define the expectation value of any functional $A[\vec{r}]$ with respect to 
$S_0$ by

\begin{eqnarray}
{\langle A[\vec{r}]\rangle }_{0}={\frac{1}{Z_{0}}}=&&\int d{\vec{r}}\int D{\vec{r}}%
(u)\exp {\{S_{0}[{\vec{r}}(u)]\}}\times  \nonumber \\
&& A[{\vec{r}}(u)]\delta({\vec{r}}(\beta )-{\vec{r}})\delta ({\vec{r}}(0)-{%
\vec{r}}).  \label{Atrial}
\end{eqnarray}

Using this trial action, the partition function of the real system, Eq. (\ref%
{Zfey}), can be written as

\begin{equation}
Z=Z_0{\langle \exp {(S-S_0)}\rangle }_0.  \label{Zave0}
\end{equation}

If the actions $S$ and $S_{0}$ are real, as is the case for our problem, the
convexity property of the function $\exp{(x)}$ can be applied from which we
find  the inequality

\begin{equation}
{\langle {\exp {(S-S_0)}}\rangle }_0\geq \exp {({\langle S-S_0\rangle }_0)}.
\label{conver}
\end{equation}
leading to a variational principle for the free energy

\begin{equation}
F\leq F_0-{\frac 1\beta }{\langle S-S_0\rangle }_0.  \label{freevari}
\end{equation}

We propose a trial action derivable from the trial Hermitian Hamiltonian

\begin{equation}
H_{0}={\frac{p_{x}^{2}}{2m}}+{\frac{p_{X}^{2}}{2M}}+{\frac{1}{2}}K(x-X)^{2}+{%
\frac{p_{y}^{2}}{2m}}+{\frac{1}{2}}m{\omega ^{\prime }}^{2}y^{2},
\label{trialH}
\end{equation}%
where $X\ (p_X)$ is the 1D position (momentum) of the fictitious particle
of mass $M$, $K$ is the strength of the harmonic potential and $\omega
^{^{\prime }}$ is the oscillation frequency associated with the lateral
confinement of the electron in the $y$ direction. The first three terms in
Eq. (\ref{trialH}) correspond to the 1-D Feynman polaron in the $x$
direction.\cite{degani86} The last two are associated to the confined motion
and the variables $M$, $K$ and $\omega ^{^{\prime }}$ are variational
parameters.

Using the same procedure as for the original Hamiltonian, Eq. (\ref{H}), we
eliminate the coordinates of the fictitious particle, and obtain the trial
action

\begin{eqnarray}
S_{0} &=&-{\frac{m}{{2}}}\int_{0}^{\beta }dt({\dot{x}(t)}^{2}+{\dot{y}(t)}%
^{2})-{\frac{1}{{2}}}m{\omega {^{\prime }}}^{2}\int_{0}^{\beta }dty(t)^{2} 
\nonumber \\
&&-{\frac{{\hbar }^{2}K^{2}}{{4M\omega }}}\int_{0}^{\beta }dt\int_{0}^{\beta
}dsG_{\omega }(t-s)(x(t)-x(s))^{2},  \label{S0trial}
\end{eqnarray}
with $\omega =\sqrt{K/M}$ and ${\it {v}}^{2}=K[\left( M+m\right) /Mm]{.}$

Following the same approach used by Peeters and Devreese \cite{peeters1},
from Eqs. (\ref{Sreal}) and (\ref{S0trial}), the Feynman variational
principle for the free energy of our system, Eq. (\ref{freevari}), can be
written as

\begin{equation}
F\leq F_F-F_R-{\frac 1\beta }{\langle S-S_0\rangle }_0,  \label{freevari1}
\end{equation}
where $F_F$ is the free energy of the trial Hamiltonian, Eq. (\ref{trialH}),
given by

\begin{mathletters}
\begin{eqnarray}
F_F &=&-{\frac 1{{\beta }}}\log [L_x{\frac vw}{\frac 1{\sqrt{2\pi \beta }}}]
\nonumber \\
&&-{\frac 1{{\beta }}}\log [\{{\frac 1{{2}\sinh {{\frac 12}\beta {\it {v}}}}}%
\}\{{\frac 1{{2\sinh {{\frac 12}\beta \omega {^{\prime }}}}}}\}],  \label{FF}
\end{eqnarray}
$F_R$ is the free energy of the fictitious particle without the interaction
term, given by

\begin{equation}
F_R=-{\frac 1{{\beta }}}\log [{\frac 1{2\sinh {{\frac 12}\beta \omega }}}],
\label{FR}
\end{equation}
and

\begin{eqnarray}
{\langle S-S_{0}\rangle }_{0}& &={\frac{1}{2}}m({\omega ^{\prime }}^{2}- {%
\omega _{0}}^{2})\int_{0}^{\beta }dt{\langle y(t)^{2}\rangle }_{0}+ \sum_{%
\vec{k}}|V_{\vec{k}}|^{2}\times  \nonumber \\
&&\int_{0}^{\beta}dt\int_{0}^{\beta} dsG_{\omega _{k}}(t-s){\langle
e^{\imath {\vec{k}}\cdot (\vec{r}(t)-\vec{r}(s))}\rangle }_{0}  \label{sms0}
\\
&&-{\frac{{\hbar m}}{4}}\omega ({\it {v}^{2}-{\omega }^{2})\int_{0}^{\beta
}dt\int_{0}^{\beta }ds\times}  \nonumber \\
&&G_{\omega }(t-s){\langle(x(t)-x(s))^{2}\rangle }.  \nonumber
\end{eqnarray}

Defining

\end{mathletters}
\begin{equation}
{\langle e^{\imath {\vec{k}}\cdot (\vec{r}(t)-\vec{r}(s))}\rangle }%
_0=A_x(k_x,t-s)A_y(k_y,t-s),  \label{expmed}
\end{equation}
where

\begin{equation}
A_\xi (k_\xi ,t-s)={\langle e^{\imath k_\xi (\xi (t)-\xi (s))}\rangle }_0
\label{Ai}
\end{equation}
with $\xi =x,y$, we have that

\begin{equation}
{\langle (\xi (t)-\xi (s))^{2}\rangle }_{0}=-{\frac{{\partial ^{2}}}{{%
\partial k_{\xi }}^{2}}}A_{\xi }(k_{\xi },t-s).  \label{xquadm}
\end{equation}

The three averages in Eq. (\ref{sms0}) can be calculated in an analogous way
as done by Osaka \cite{osaka} and Peeters and Devresee,\cite{peeters1} such
that we obtain

\begin{equation}
{\frac 1m}{\langle y(t)^2\rangle }_0={\frac 1{{2\omega {^{\prime }}}}}\coth {%
\frac{\beta \omega {^{\prime }}}2},  \label{y2}
\end{equation}

\begin{equation}
A_\xi (k_\xi ,t-s)=e^{-{k_\xi }^2D_\xi (t-s)}  \label{axi}
\end{equation}
with

\begin{mathletters}
\begin{eqnarray}
D_x(t) &=&{\frac{|t|{\omega }^2}{{2{\it v}^2}}}(1-{\frac{|t|}{{\beta }}}) 
\nonumber \\
&&-{\frac{{{{\it v}^2-{\omega }^2}}}{{2{\it v}^3}}}(1-e^{-{\it v}|t|}-4n(%
{\it v})\sinh ^2{\frac{{\it v}t}2}),  \label{dx}
\end{eqnarray}
and $D_y(t)$ is obtained from the above expression by taking $\omega
\rightarrow 0$ and ${\it v\rightarrow \omega {^{\prime }}}$, giving

\begin{equation}
D_y (t) = {\frac{1 }{{2 \omega{^{\prime}}}}}(1- e^{-{\omega{^{\prime}}}|t|}
-4n({\it v}) \sinh^2{\frac{{\omega{^{\prime}}}t }{2}})\quad .  \label{dy}
\end{equation}

Substituting Eqs. (\ref{FF})-(\ref{dy}) in Eq. (\ref{freevari1}), we finally
obtain an upper bound $F^v$ for the exact free energy of our system, given by

\end{mathletters}
\begin{eqnarray}
F^{v} &&=\frac{1}{4{\it v}}({\it v}^{2}-{\omega }^{2})\left(\coth\frac{\beta%
{\it v}}{2}-\frac{2}{\beta{\it v}}\right )-\frac{1}{\beta}\log\left[\frac{v}{%
w}\frac{L_x}{\sqrt{2\pi \beta }}\right]  \nonumber \\
&&-\frac{1}{\beta} \log\left[\left( \frac{\sinh{\frac{1}{2}\beta \omega }}{%
\sinh{\frac{1}{2}\beta {\it v}}}\right) \left\{ \frac{1}{{2\sinh {{\frac{1}{2%
}} \beta \omega {^{\prime }}}}}\right\}\right]-\frac{1}{4{\omega^{\prime}}}%
\times  \nonumber \\
&&({\omega^{\prime}}^{2}-{\omega _{0}}^{2}) \coth{\frac{\beta\omega^{\prime }%
}{2}}-\sum_{\vec{k}}|V_{\vec{k}}|^{2} [1+n({\omega}_{\vec{k}})]\times
\label{Fv} \\
&&\int_{0}^{\beta} du e^{-{\omega}_{\vec{k}}u}e^{-{k_{x}}^{2}D_{x}(u)}e^{-{%
k_{y}}^{2}D_{y}(u)},  \nonumber
\end{eqnarray}
which is subject to the minimization conditions

\begin{equation}
{\frac{\partial F^v}{\partial {\it v}}}={\frac{\partial F^v}{\partial {%
\omega }}}={\frac{\partial F^v}{\partial {\omega {^{\prime }}}}}=0.
\label{varia}
\end{equation}

\section{GROUND-STATE ENERGY AND LIMITING CASES}

The result obtained in Eq. (\ref{Fv}) is general and can be used to
calculate all thermodynamic quantities, like e.g. specific heat, entropy,
internal energy, etc. In the zero-temperature limit, the free energy $F^v$
reduces to the polaron ground-state energy ${E_0}^v$, given by

\begin{eqnarray}
{E_{0}}^{v}&&=\lim_{\beta \rightarrow \infty }F^{v}= {\frac{1}{2}}\omega {%
^{\prime }}-{\frac{({\omega ^{\prime }}^{2}-{\omega _{0}}^{2})}{{4{\omega {%
^{\prime }}}}}}+ {\frac{1}{{4{\it v}}}}({\it v}-{\omega })^{2}-  \nonumber \\
&&\sum_{\vec{k}}|V_{\vec{k}}|^{2}\int_{0}^{\infty }due^{-{\omega }_{\vec{k}%
}u}e^{-{k_{x}}^{2}{D_{x}}^{0}(u)}e^{-{k_{y}}^{2}{D_{y}}^{0}(u)},  \label{E0v}
\end{eqnarray}
with

\begin{mathletters}
\begin{equation}
{D_x}^0(u)={\frac{{\omega }^2}{{2{\it v}^2}}}u+{\frac{{{{\it v}^2-{\omega }^2%
}}}{{2{\it v}^3}}}(1-e^{-{\it v}u}),  \label{dx0}
\end{equation}
and $D_y(t)$ is obtained from the above expression by taking $\omega
\rightarrow 0$ and ${\it v\rightarrow \omega {^{\prime }}}$, giving

\begin{equation}
{D_y}^0 (u) = {\frac{1 }{{2 \omega{^{\prime}}}}}(1- e^{-{\omega{^{\prime}}}%
u}).  \label{dy0}
\end{equation}

In Eq. (\ref{E0v}) the first term corresponds to the zero-point energy of
the harmonic oscillator with frequency $\omega ^{^{\prime }}$, the second
one is due to the interaction between the lateral confinement and
the fictitious oscillator with frequencies $\omega _{0}$ and $\omega
^{^{\prime }}$, respectively. The third term is due to the Feynman polaron
with eigenfrequency $v$ and the fictitious particle with frequency $w$ (i.e. 
$m\rightarrow \infty $). The last term is associated with the
electron-phonon interaction.

If the $y$-confinement is much larger than the phonon effects, one can take
the limit $\omega {^{\prime }}\approx \omega _0$. Also, if in this limit the
phonon cloud surrounding the electron is very small, we can assume $v\approx
w$. Within these approximations, the polaron ground-state energy ${E_0}^v$,
Eq.(\ref{E0v}), is given by

\end{mathletters}
\begin{eqnarray}
{E_{0}}^{conf}&=&{\frac{1}{2}}\omega _{0}-\sum_{\vec{k}}|V_{\vec{k}%
}|^{2}\times  \nonumber \\
&&\int_{0}^{\infty }due^{-{\omega }_{\vec{k}}u}e^{-{k_{x}}^{2}u/2}e^{-{\
k_{y}}^{2}(1-e^{-\omega _{0}u})/{2\omega _{0}}},  \label{E0vapp0}
\end{eqnarray}
which equals the result from second-order perturbation theory.

The first term in ${E_{0}}^{conf}$ is the zero-point energy of the harmonic
oscillator, the second exponential term under the integral is associate with
the free motion along the $x$ direction, and the last exponential under the
integral is due to the oscillator motion in the confinement direction
containing all confinement levels. Expanding the last exponential term under
the integral of ${E_{0}}^{conf}$, we can integrate out the $u$ variable
resulting in

\begin{eqnarray}
{E_{0}}^{conf}&=&{\frac{1}{2}}\omega _{0}-\sum_{\vec{k}}|V_{\vec{k}%
}|^{2}\times  \nonumber \\
&&\sum_{n=0}^{\infty }{\frac{1}{{n!}}}\left({\frac{{k_{y}}^{2}}{{2\omega _{0}%
}}}\right)^{n}{\frac{e^{-{k_{y}}^{2}/{2\omega _{0}}}}{{\omega
_{k}+k_{x}^{2}/2+n\omega _{0}}}},  \label{E0vapp}
\end{eqnarray}
which allows a perturbation analysis in terms of diagrams.

We now consider the limiting cases of weak and strong coupling. For weak
coupling, we have the following conditions: $\alpha \gg 0.5$, $v\approx w$,
and $\omega ^{\prime }\rightarrow 0$. Therefore, expanding the exponentials
in the $D_{x,y}$ terms, we obtain $D_{x}(u)\approx u/2$ and $D_{y}(u)\approx u/2
$, resulting in $f(k,u)\simeq \exp (-k^{2}u/2)$. In this case, the integral

\begin{eqnarray}
B&=&\alpha \int_{0}^{k_{0}}dkk^{2}\int_{0}^{\infty}due^{-ku}f(k,u)  \nonumber
\\
&\simeq & 2\alpha k_{0}\left[1-\frac{2}{k_{0}}ln\left(1+\frac{k_{0}}{2}%
\right)\right],
\end{eqnarray}
and the polaron energy can be written as 
\begin{equation}
E_{pol}=\frac{1}{4}\left( \frac{\omega _{0}^{2}}{\omega ^{\prime }}\right)
-B.
\end{equation}%
In the absence of confinement ($\omega _{0}=0),$ even when $\omega ^{\prime
} $ is too small, $E_{pol}=-B$. For $\omega _{0}>1,$ the polaron energy
increases fast.

In the strong limit, $\alpha \gg 0.5,$ we have the conditions: $v\gg w$, and 
$\omega ^{\prime }\rightarrow v$. Therefore, as before, expanding the
exponentials in the $D_{x,y}$ terms we obtain now $D_{x}(u)\approx 1/2v$ and 
$D_{y}(u)\approx 1/2\omega ^{\prime }$, resulting in $f(k,u)\simeq \exp
(-k^{2}/2v).$ In this case, the integral 
\begin{eqnarray}
B&=&\alpha \int_{0}^{k_{0}}dkk^{2}\int_{0}^{\infty }due^{-ku}f(k,u) 
\nonumber \\
&\simeq&-\alpha v\left( e^{-k_{0}^{2}/2v}-1\right),
\end{eqnarray}
and the energy is given as 
\begin{equation}
E_{pol}=\frac{v}{2}-\frac{\omega _{0}^{2}}{4v}+\alpha v\left(
e^{-k_{0}^{2}/2v}-1\right) .
\end{equation}%
One can see that the energy decreases when we increase the lateral
confinement.

\section{NUMERICAL RESULTS}

In this section we present the numerical results for the zero-temperature
limit of the free energy $F^{v}$, that is, the confinement polaron
ground-state  energy ${E_{0}}^{v}$, by minimizing Eq.(\ref{E0v}), and
considering the  electron interacting with longitudinal surface acoustic
phonons (ripplons).  For these excitations, the dispersion relation is given
by

\begin{mathletters}
\begin{equation}
\omega _{\vec{k}}=s|\vec{k}|,  \label{disp}
\end{equation}
where $s$ is the velocity of sound, and the Fourier transform of the
electron-phonon interaction

\begin{equation}
\left| V_{\vec{k}}\right| ^{2}\propto \alpha k,  \label{poten}
\end{equation}
describes both 2D phonons and ripplons\cite{polaron,farias96}, where $\alpha 
$ is the dimensionless electron-phonon coupling constant which depends on
the deformation potential. In our calculation the sum over the phonon wave
vectors $\sum_{\vec{k}}$ will be replaced by the integral $A/(2\pi )^{2}\int
d\vec{k}$, which is cutoff at $k_{0}$, the Debye critical wave vector in
phonon space which simulates the discreteness of the lattice, $k_{0}\sim
1/a, $ with $a$ the lattice constant, and corresponds to the capillary
constant $k_{c}\sim 1/d^{2}$, where $d$ is the film thickness in case of
liquid helium. Hereafter, we use dimensionless units and express the energy
in units of $ms^{2}$ and the length in units of $\hbar /ms$. Substituting
Eqs. (\ref{disp})-(\ref{poten}) in Eq. (\ref{E0v}), and integrating out the
angular coordinate of $\vec{k}$ we obtain

\end{mathletters}
\begin{eqnarray}
{E_{0}}^{v} &=&{\frac{1}{2}}\omega {^{\prime }}-{\frac{1}{{4{\omega {%
^{\prime }}}}}}({\omega {^{\prime }}}^{2}-{\omega _{0}}^{2})+{\frac{1}{{4%
{\it v}}}}({\it v}-{\omega })^{2}  \nonumber \\
&&-\alpha \int_{0}^{k_{0}}dkk^{2}\int_{0}^{\infty }due^{-ku}f(k,u),
\label{Eacoust}
\end{eqnarray}
with

\begin{eqnarray}
f(k,u)=&&e^{-{{\frac 12}{k_x}^2({D_x}^0(u)+{D_y}^0(u))}}\times  \nonumber \\
&&I_0\left({\frac{{{k_x}^2}}2}|{{D_x}^0(u)-{D_y}^0(u)}|\right).  \label{fku}
\end{eqnarray}

The results for the energy as a function of $\alpha $ are shown in Fig.\ 1,
for different values of the Debye cutoff $k_{0}$ and the lateral confinement
frequency $\omega _{0}$. The figure shows that the slope of the polaron
energy varies at $\alpha \approx 0.51$. This value of $\alpha$ is
approximatelly twice the value obtained in Ref.10, even when we take the
limit $\omega _{0}\rightarrow 0$. This difference can be understood as due 
to the fact that the translational invariance in the $y$ direction is not
taken into account in our model Hamiltonian, i.e, we are considering a Q1D
system.

As can be seen, the lateral confinement increases the energy and this is
more significant for small values of $\alpha $. In order to analyze the
effects of the lateral confinement on the self-trapping transition, Fig.\ 2
shows the first derivative of the ground-state energy as a function of $%
\alpha $ for two values of $k_{0}$. We observe, in Fig.\ 2(a), that the
lateral confinement, $\omega _{0}=20$, smoothens the curve compared with the
one where $\omega _{0}=0$. Considering $k_{0}=150$, Fig.\ 2(b) shows the
existence of a discontinuity in the first derivative corresponding to a
first-order transition in the case where $\omega _{0}=0$. This is not
observed for $\omega _{0}=20$. These facts are best visualized in the second
derivative of the ground-state energy as function of the coupling constant
presented in Fig.\ 3. 

In Fig. 4, the phase diagram ($k_{0},\omega _{0})$ for the 2D acoustical
polaron is depicted for $\alpha =0.515$ where the transition occurs. For
each $\omega _{0}$, the value of $k_{0}$ determines the self-trapped
transition and the curve bounds two regions. In the upper region, the
transition is discontinuous, while in the below region it is continuous
for this $\alpha$-value. In the latter region it is possible to
find a discontinous self-trapping transition when $\alpha >0.515.$

In Fig.\ 5 we present the Feynman polaron mass $M=(v/w)^{2}$ as a function
of $\alpha $, for different values of the Debye cutoff $k_{0}$ and the
lateral confinement frequency $\omega _{0}$. The figure shows clearly the
transition from the weak- to strong-coupling regime. It also shows that for $%
k_0<70$ the transition is continuous. Again, it is worth to mention that,
for $\omega _{0}=0.0$, the value obtained here is different from the the 2D
polaron.\cite{farias96} Once more, this difference is due to the fact that
the translational invariance in the $y$ direction is not taken into account
in our model  Hamiltonian.

We also observe that the effects of the lateral confinement on the polaron
mass are more significant in the region of small $\alpha $, i.e. the non
self-trapped region, and the transition from low to large polaron mass is
now smooth.

In Fig.\ 6 we show the ground-state energy shift $\Delta E=E(\omega
_{0}=0)-E(\omega _{0})$ as a function of $\omega _{0}$, for different values
of $k_{0}$ and a fixed value of $\alpha $. The lateral confinement increases
the binding energy and we do not see any abrupt change in the energy values.
This fact can be understood from Eq. (\ref{E0vapp}). This result shows that,
when $\omega _{0}$ increases, the zero-point energy increases more than the
terms corresponding to the free motion along the $x$ direction, and the one
associated to the oscillator motion in the confinement direction, that
contain all the confinement levels. Although the ground-state energy does
not present an abrupt change with the confinement this fact is observed in
the Feynman polaron mass. In Fig.\ 7 we present the Feynman polaron mass $%
M=(v/w)^{2}$ as a function of $\omega _{0}$, for different values of $k_{0}$
and a fixed value of $\alpha $. As can be seen, the polaron mass increases
with the lateral confinement and can change dramatically for large values of 
$k_{0}$ where it exhibits a discontinuous behavior. 

\section{CONCLUSIONS}

In this work we applied the Feynman path-integral method to study the
effects of lateral confinement on the ground-state properties of the 2D
acoustical polaron which can be formed in Q1D electron systems over a
helium film. We evaluated the ground-state energy and the polaron mass as a
function of the electron-phonon coupling $\alpha $ and we analyzed the
effect of lateral confinement on these properties. We determined the phase
diagram for polaron formation and found the intriguing result that the
self-trapping transition becomes more continuous when the electron motion is
reduced from 2D to a Q1D system.

\acknowledgments G.A.F, R.N.C.F and N.S are supported by the Conselho
Nacional de Desenvolvimento Cient\'{\i}fico e Tecnol\'{o}gico (CNPq) and
F.M.P is sponsored by the Belgian National Science Foundation
and the EC-project: HPRN-CT-2000-00157. N.S is
grateful to Funda\c{c}\~{a}o de Amparo \`{a} Pesquisa do Estado de S\~{a}o
Paulo (FAPESP) for a research grant.

\begin{center}
\vspace{0in}\newpage FIGURES
\end{center}

\noindent Fig. 1. Ground-state energy of the 2D acoustical polaron as a
function of the electron-phonon coupling constant $\alpha $ for different
values of $k_{0}$, $\omega _{0}=20$ (dash-dotted lines) and $\omega _{0}=0$
(solid lines).

\medskip

\noindent Fig. 2. First derivative of the ground-state energy of the 2D
acoustical  polaron as a function of the electron-phonon coupling constant $%
\alpha $ for: a) $k_{0}=50,$ b) $k_{0}=150,$ in both cases for $\omega_{0}=0$%
(solid lines) and $\omega _{0}=20$ (dash-dotted lines).

\medskip

\noindent Fig. 3. Second derivative of the ground-state energy of the 2D
acoustical  polaron as a function of the electron-phonon coupling constant $%
\alpha $ for: a) $k_{0}=50,$ and b) $k_{0}=150,$ in both cases for $%
\omega_{0}=0$  (solid lines) and $\omega _{0}=20$ (dash-dotted lines).

\medskip

\noindent Fig. 4. Phase diagram ($k_{0},\omega _{0}$) for the self-trapping
transition of the acoustical polaron at $\alpha =0.515$.

\medskip

\noindent Fig. 5. Feynman polaron mass $M=(v/w)^{2}$ as a function of the
electron-phonon coupling constant $\alpha $ for different values of $k_{0}$
and $\omega _{0}.$

\medskip 

\noindent Fig. 6. The ground-state energy shift of the polaron, $\Delta
E=E(\omega _{0}=0)-E(\omega _{0})$ as a function of the lateral confinement
frequency $\omega _{0}$ for $\alpha =0.515$, considering: $k_{0}=50$ the
solid line, $k_{0}=100$ the dashed line, and $k_{0}=100$ the dash-dotted
line.

\medskip

\noindent Fig. 7. Polaron mass as a function of the lateral confinement
frequency $\omega _{0}$for the same parameters as in Fig. 6.

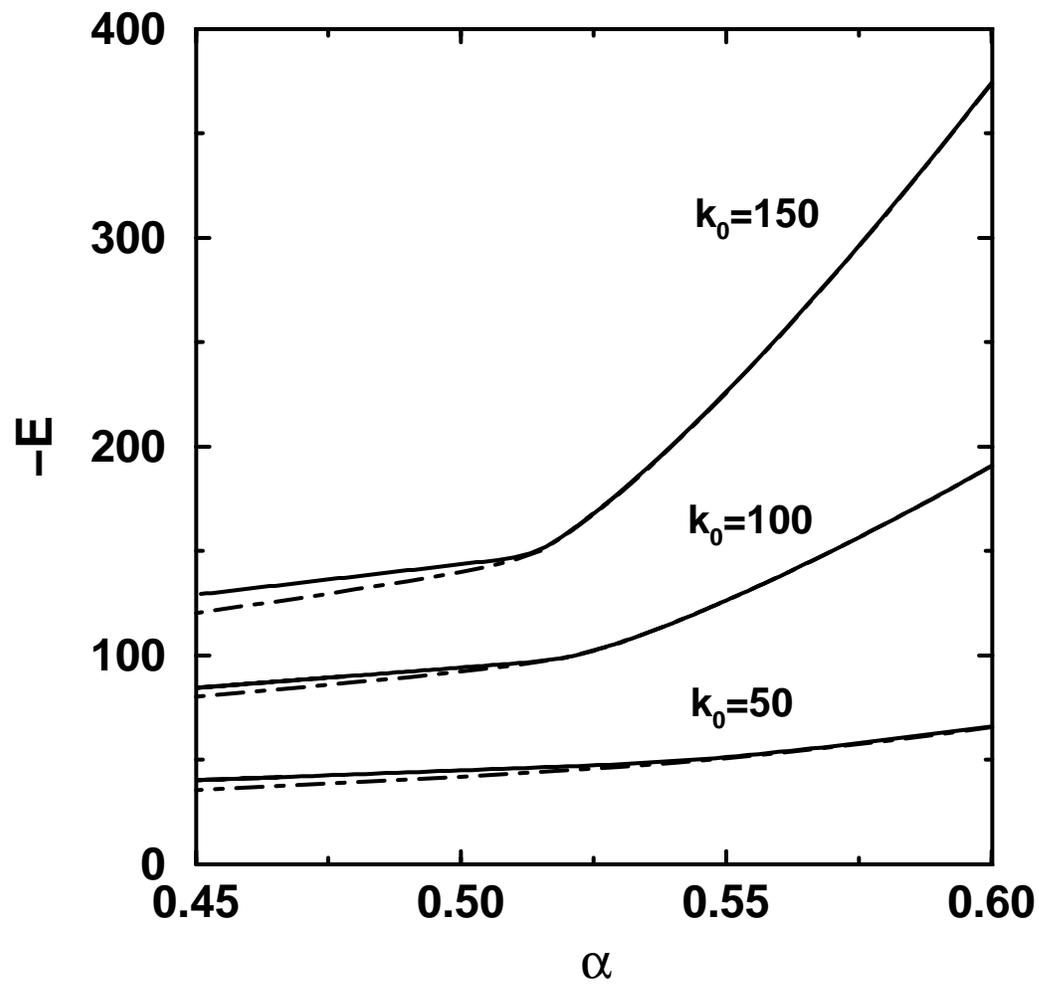
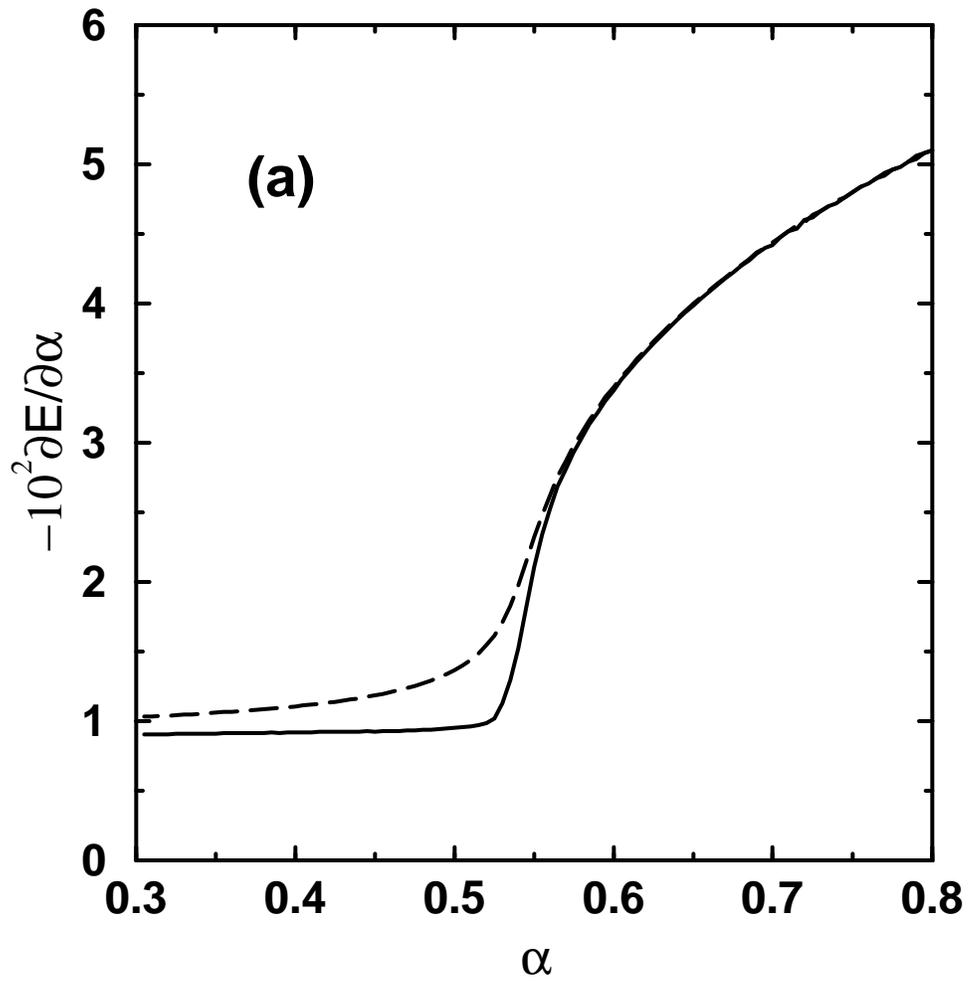
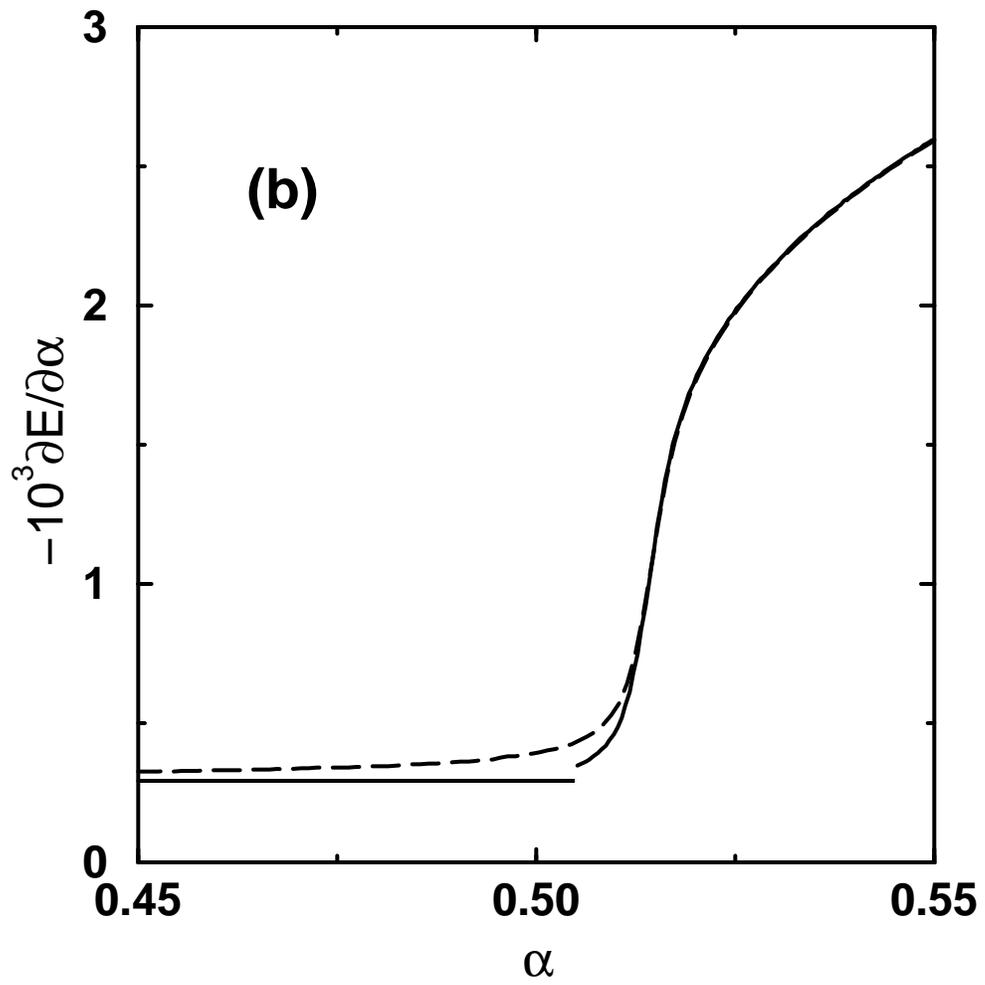
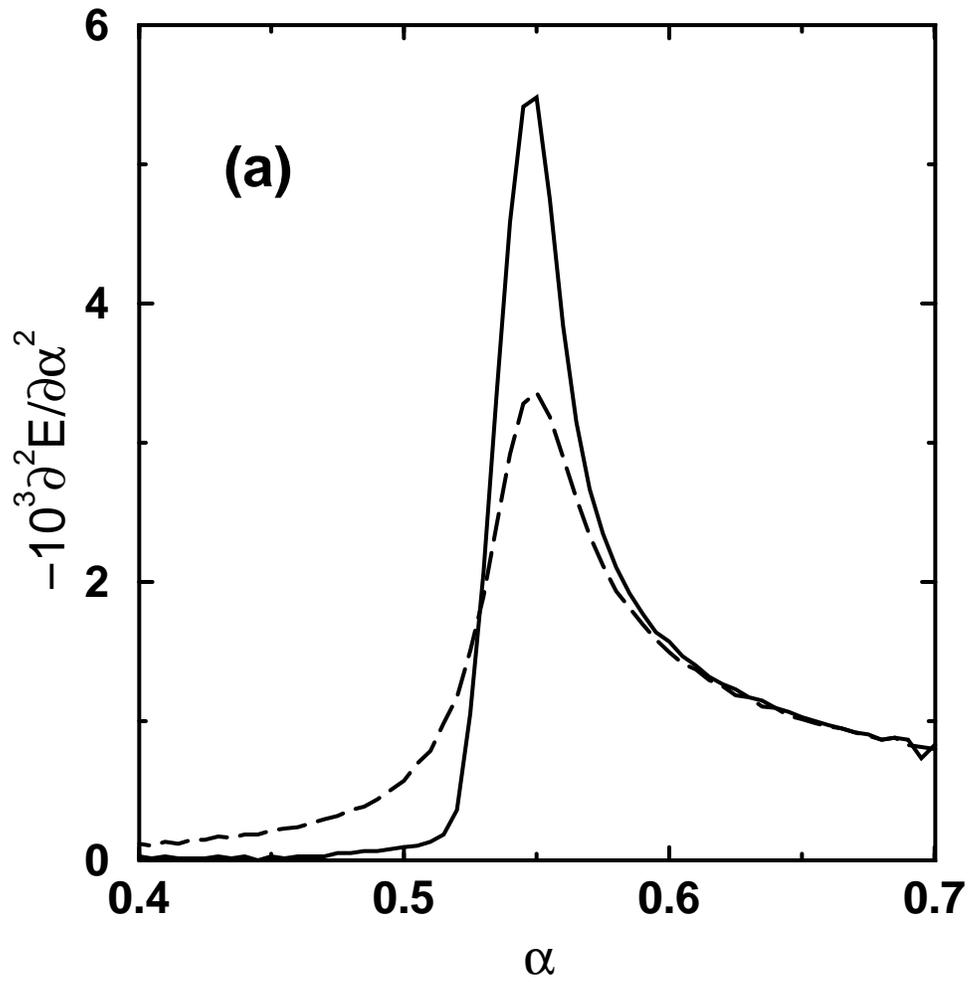
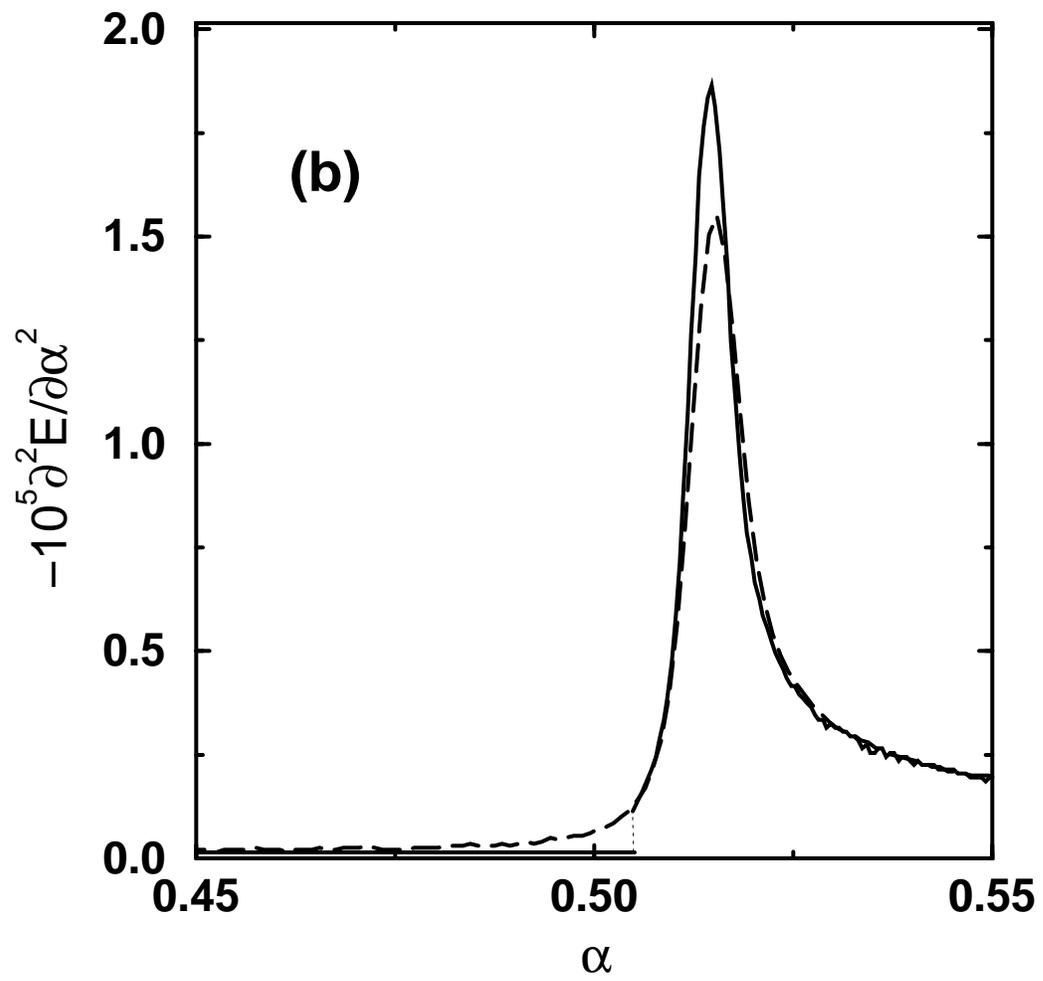
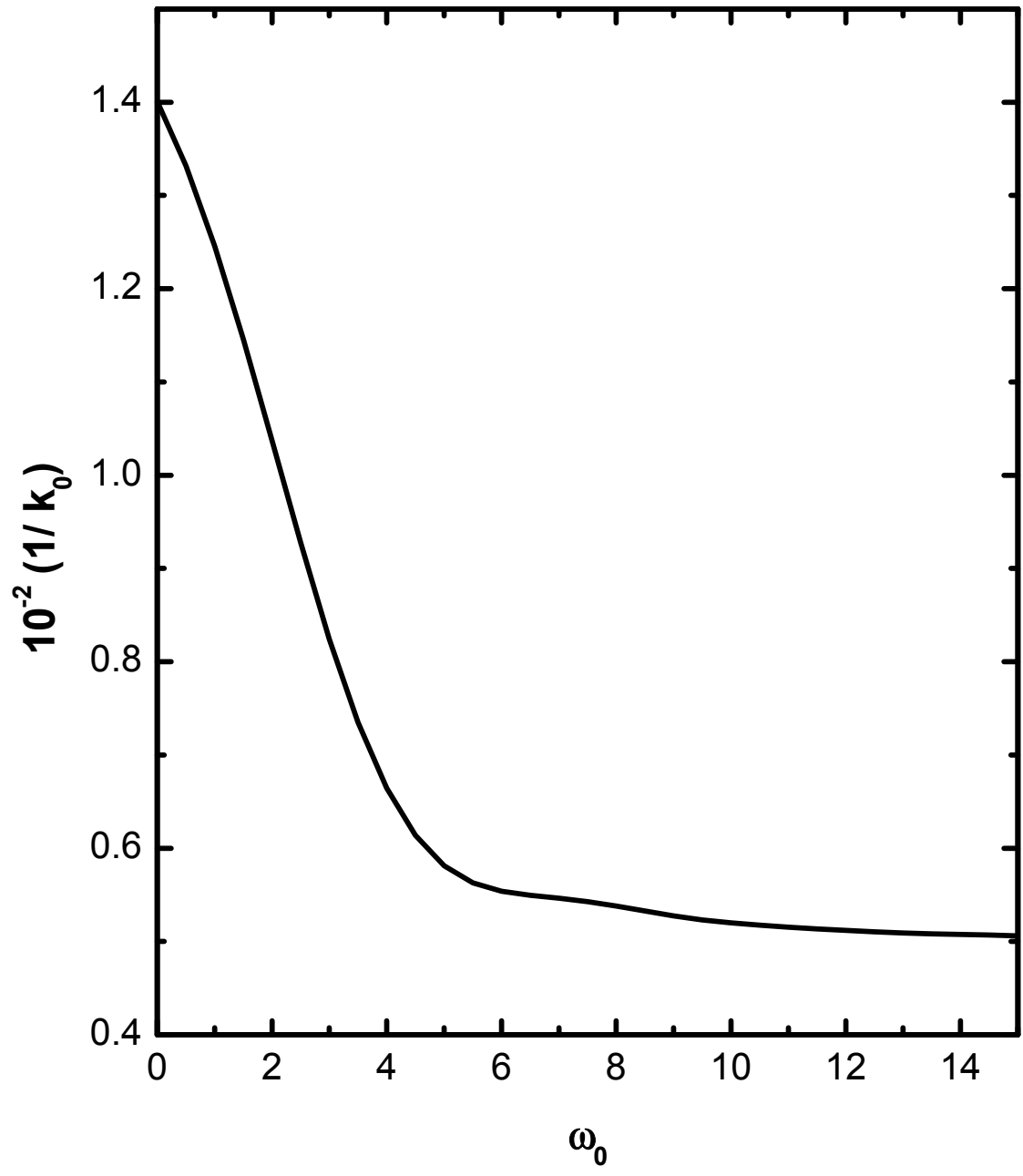
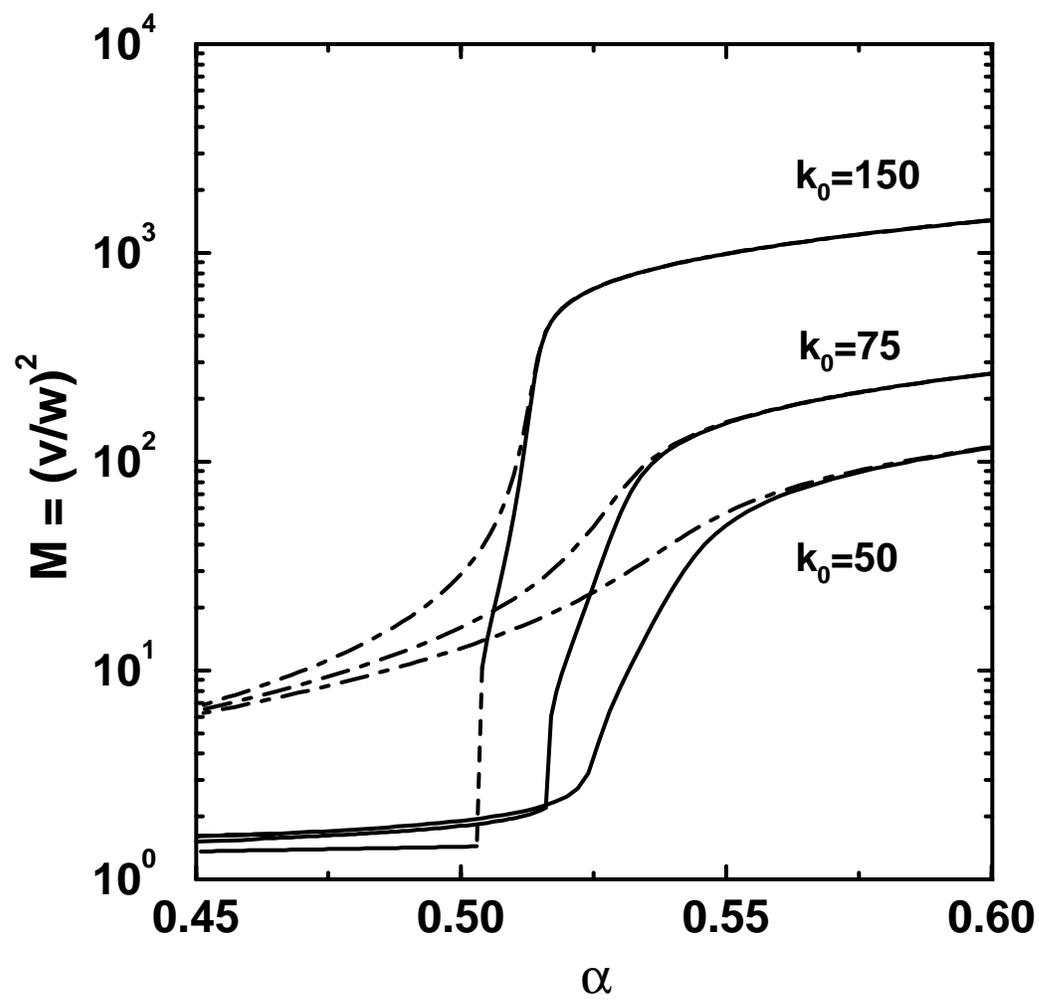
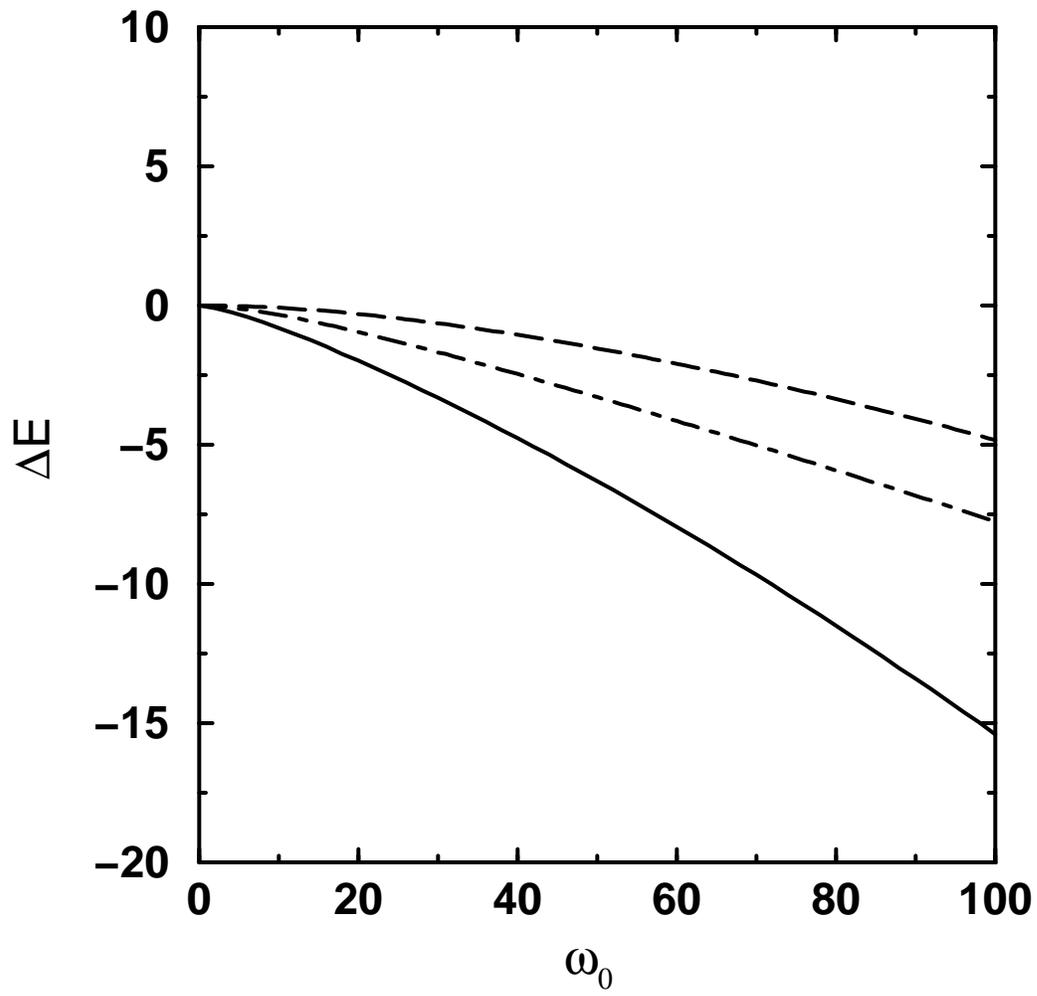
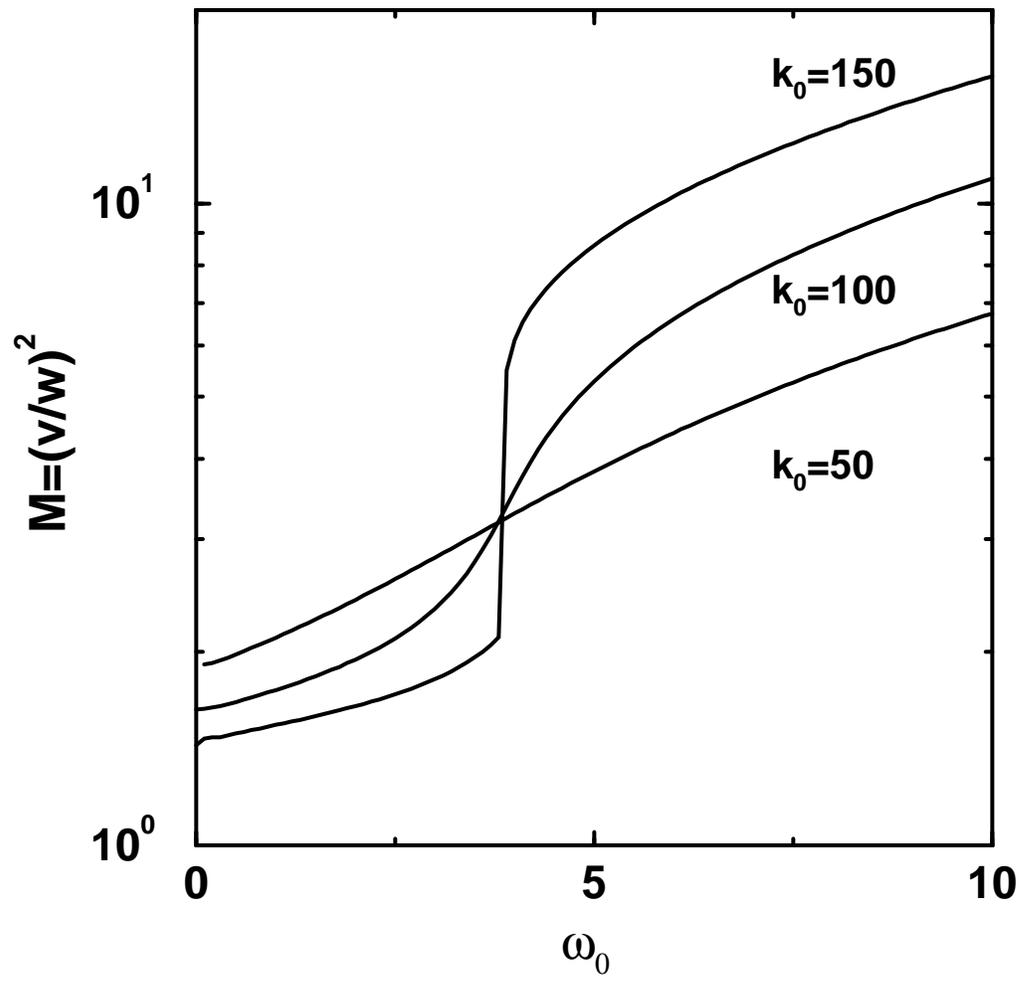
\end{document}